\documentclass[aps,prb,twocolumn,floatfix,footinbib,showpacs,superscriptaddress]{revtex4-1}
\usepackage{graphicx}
\usepackage{amsfonts,amsmath,amssymb}
\usepackage{amsthm}
\usepackage{dsfont,bm}
\usepackage{color}
\usepackage{soul} 
\usepackage{amsbsy}
\usepackage[colorlinks=true,linkcolor=blue,pagecolor=blue,filecolor=blue,menucolor=blue,urlcolor=blue,citecolor=blue,anchorcolor=blue]{hyperref}%
\usepackage{sidecap}
\usepackage{txfonts}
\usepackage{amsbsy}
\usepackage{dcolumn}

\newcommand{\dr}{d{\bm r}}

\newcommand{\ua}{\uparrow}
\newcommand{\da}{\downarrow}

\begin{document}

\title{Strain-engineered Majorana Zero Energy Modes and $\varphi_0$ Josephson State in Black Phosphorus }

\author{Mohammad Alidoust}
\affiliation{Department of Physics, K.N. Toosi University of Technology, Tehran 15875-4416, Iran}
\author{Morten Willatzen}
\affiliation{Beijing Institute of Nanoenergy and Nanosystems, Chinese Academy of Sciences, No. 30 Xueyuan Road, Haidian District, Beijing 100083, China}
\affiliation{Department of Photonics Engineering, Technical University of Denmark, DK-2800 Kongens Lyngby, Denmark}
\author{Antti-Pekka Jauho}
\affiliation{Center for Nanostructured Graphene (CNG), Department of Micro- and Nanotechnology, Technical University of Denmark, DK-2800 Kongens Lyngby, Denmark}

\date{\today}
\begin{abstract}
We develop a theory for strain control of Majorana zero energy modes and the Josephson effect in black phosphorus (BP) devices proximity coupled to a superconductor. Employing realistic values for the band parameters subject to strain, we show that the strain closes the intrinsic band gap of BP; however, the proximity effect from the superconductor reopens it and creates Dirac and Weyl nodes. Our results illustrate that Majorana zero energy flat bands connect the nodes within the band-inverted regime in which their associated density of states is localized at the edges of the device. In a ferromagnetically mediated Josephson configuration, the exchange field induces superharmonics in the supercurrent phase relation in addition to a $\varphi_0$ phase shift, corresponding to a spontaneous supercurrent, and strain offers an efficient tool to control these phenomena. We analyze the experimental implications of our findings and show that they can pave the way for creating a rich platform for studying two-dimensional Dirac and Weyl superconductivity.
\end{abstract}
\maketitle

\section{introduction}

The topological phases in condensed matter have recently attracted robust attention both theoretically and experimentally due to the unique properties they offer \cite{kane_rmp,zhang_rmp,Beenakker2013ARCM,Nayak2008RMP}. The topological phase supports gapless surface states which are anticipated to serve as backscattering-free channels \cite{kane_rmp,zhang_rmp}. In the presence of superconductivity, these surface modes can host Majorana fermions with zero energy [the so-called topological superconducting (TS) phase]. The Majorana fermions are their own antiparticles and are governed by non-Abelian statistics. These fermions are expected to play a key role for fault-tolerant topological quantum computation \cite{Beenakker2013ARCM,Nayak2008RMP}.

One of the main challenges in the context of topological superconductivity is that natural materials rarely host this phase \cite{Beenakker2013ARCM}. A prominent example of natural topological superconductors is the transition metal oxide superconductor ${\rm Sr_2RuO_4}$, supporting exotic pairings (such as the spin-triplet chiral $p$-wave) \cite{Maeno,Mackenzie}. Nevertheless, conclusive identification of the TS phase even in ${\rm Sr_2RuO_4}$ is still elusive \cite{Liu,Zhang1}. Topological superconductivity can also exist in Dirac and Weyl semimetals as well as in other contexts that support zero energy flat bands associated with the surface-localized Majorana zero energy modes \cite{law,OBrien,Gregoras,Li,Hou,Xiao,Alidoust,Sun,Das,Huang,Izumida,Zhang,Lv,Bachmann,Oudah,He}. The Majorana flat bands were also discussed in connection with superfluids \cite{volov1,volov2}.
Other ways to access the TS phase have also been examined. In one attempt a spin-orbit-coupled semiconductor nanowire was deposited on top of a singlet superconductor, and an external magnetic field was applied perpendicular to the nanowire \cite{Chang,Kjaergaard,Nichele,Deng,Mourik}, following theoretical suggestions \cite{Fu,Sau,Lutchyn}. The experimental observations showed zero energy peaks through tunneling experiments, and this zero-bias mode was attributed to the presence of Majorana fermions, although not conclusively \cite{Chang,Kjaergaard,Nichele,Deng,Mourik,dsarma,allgpeople,ramon}. There are also other ongoing efforts using, for example, thermal Hall conductance to observe the Majorana fermions \cite{Kasahara,Banerjee,Shapiro}. The experiments revealed fractional conductance at a filling factor of $5/2$ in quantum Hall states, providing a signature of non-Abelian states.

Despite extensive efforts during the past decade, the detection of Majorana fermions has brought up numerous experimental challenges. In addition to difficulties regarding material parameters, one main challenge is the unique interpretation of the performed experiments. For example, {\rm Fe} atoms deposited on top of a {\rm Pb} superconductor self-recombine into straight chains with a length of $\sim 50$~nm, and the measured density of states \cite{Perge} near the ends of the chains shows a mode at zero bias which is attributed to the presence of Majorana fermions. This experiment was repeated by other groups as well \cite{Pawlak,Ruby}. However, a similar peak in the density of states near the ends of such {\rm Fe} chains was also observed for ``nonsuperconducting'' substrates \cite{Menzel}. Furthermore, a recent experiment observed that the zero energy peak at the ends of chains is not well localized and indeed splits into two peaks located on lateral sides of the chain \cite{Jeon}. These difficulties underscore the fundamental importance to have at least more reliable and controlled platforms when attempting to detect the TS phase.

In order to address this need, we propose device concepts based on monolayer black phosphorus (BP) coupled to a superconductor, with an externally exerted mechanical strain as a control parameter. One advantage of BP is its intrinsic large direct band gap. Also, BP is mechanically flexible and can sustain strong strains without any rupture. Additionally, its band gap is largely tunable by an externally applied strain \cite{Carvalho,kim1}. A recent angle-resolved photoemission spectroscopy experiment on BP demonstrated a large band inversion of the order of $\sim 0.6$ eV accompanied by the creation of stable Dirac points \cite{kim2}. The two devices considered in this work are depicted in Fig. \ref{fig1}. We first state our main results and discuss the technical calculations below. In Fig. \ref{fig1}(a) a single $s$-wave superconductor (S) of length $d_S$ is proximity coupled to a BP sheet, and an external mechanical force exerts strain in the plane of the BP. Using band structure analyses with realistic band parameters, our calculations reveal that the mechanical strain is able to drive a phase transition and closes the band gap of BP, while the presence of a superconducting gap reopens it. Simultaneously, Dirac and/or Weyl nodes are created; the nodes are connected by dispersionless flat bands at zero energy. In addition, connecting single lines 
at finite energies appear. The associated densities of states are localized at the edges of the device. In the second device a ferromagnetic tunnel barrier is inserted between two superconductors with a nonzero phase difference, all proximity coupled to the BP sheet, as shown in Fig. \ref{fig1}(b). Our results for the supercurrent flowing in the BP demonstrate a $\varphi_0$ state that can be controlled by an external magnetic field, or the exerted strain. Also, the strain can change the direction of the supercurrent, suggesting an experimentally controllable switch between $0$ and $\pi$ states. Our findings offer a glimpse of a rich band engineering potential and suggest a simple platform to investigate two-dimensional Weyl and Dirac superconductivity.

\section{results and discussion}

\begin{figure}[t]
\includegraphics[clip, trim=2.5cm 3.7cm 2.7cm 0.8cm, width=7.50cm,height=5.50cm]{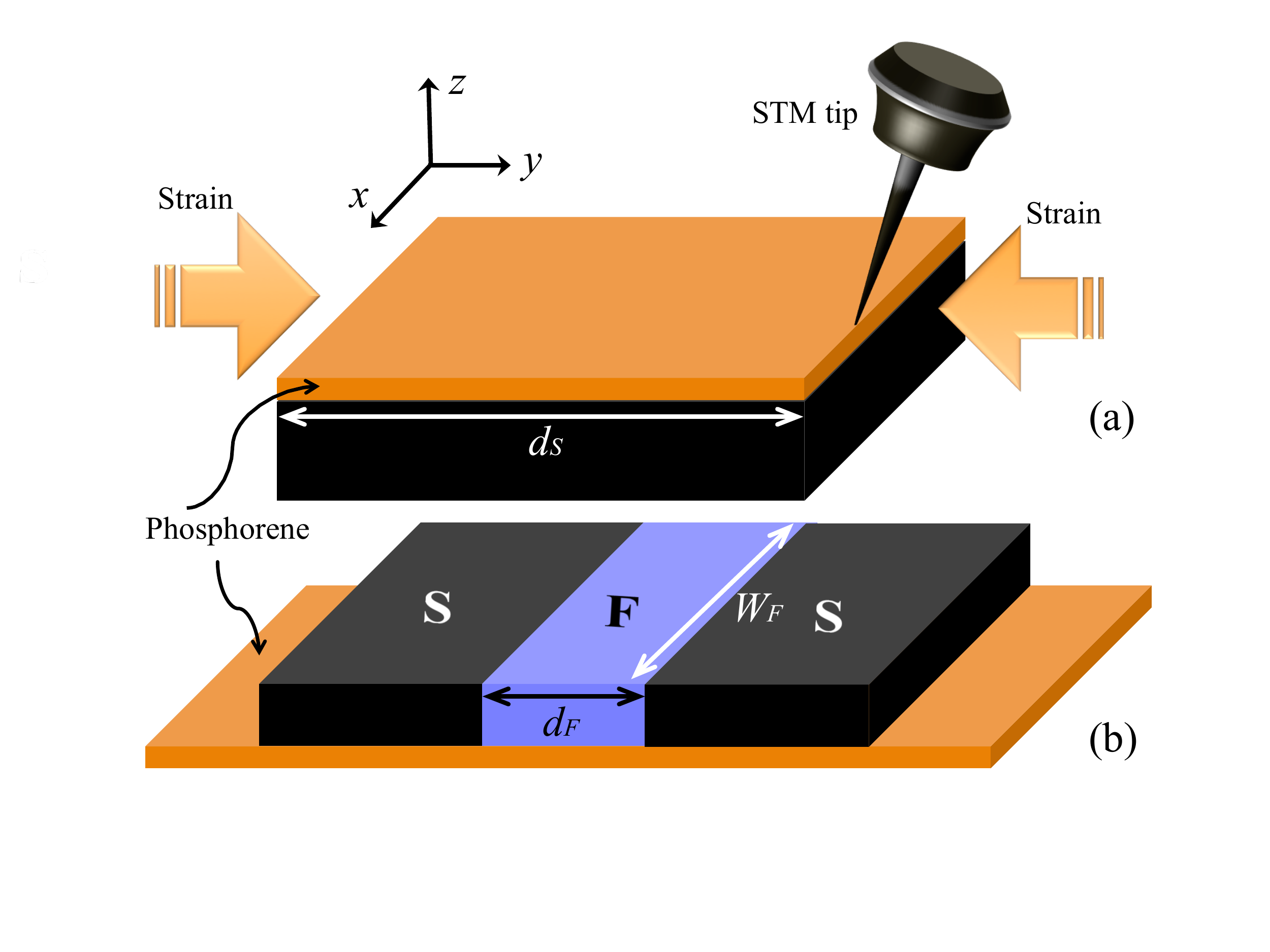}
\caption{\label{fig1} (Color online).
Schematic of the proposed devices made of a phosphorene monolayer. The plane of black phosphorus is located in the $xy$ plane, and strain is exerted into the plane of the setups in the $y$ direction. (a) A single $s$-wave superconducting electrode (S) of length $d_S$ is proximity coupled to black phosphorus. In order to reveal the density of states in the vicinity of the junction location along the $x$ axis, the STM tip should be placed close to the superconductor edge and black phosphorus. (b) A ferromagnetic (F) phosphorene Josephson junction of length $d_F$ and width $W_F$.}
\end{figure}

To model the devices displayed in Figs. \ref{fig1}(a) and \ref{fig1}(b), we employ the following low energy Hamiltonian describing a monolayer BP with strain tensor components $\varepsilon_{ii}$: 
\begin{eqnarray}
H= &&\int \frac{d\textbf{k}}{(2\pi)^2}\hat{\psi}^\dag_{\textbf{k}}H(\textbf{k})\hat{\psi}_{\textbf{k}}=\nonumber\\ &&\int \frac{d\textbf{k}}{(2\pi)^2}\hat{\psi}^\dag_{\textbf{k}}\Big\{ \big [u_0+\alpha_i\varepsilon_{ii} +(\eta_j+\beta_{ij}\varepsilon_{ii})k_j^2\big]\tau_0 +\nonumber\\ &&\big[\delta_0+\mu_i\varepsilon_{ii} +(\gamma_j+\nu_{ij}\varepsilon_{ii})k_j^2\big ]\tau_x -\chi_y k_y\tau_y\Big\}\hat{\psi}_{\textbf{k}} 
\end{eqnarray}
where indices ($i,j$) run over coordinates $x,y$. Here $\tau_i$ are the Pauli matrices in the pseudospin space, and the vector quantities are denoted by boldface. In the presence of magnetism, ${\bm h}=(h_x,h_y,h_z)$, we invoke the real spin space and attach three directions of spin to each valley, denoted by $h_x\sigma_{x}+ h_y\sigma_{y}+ h_z\sigma_{z}$, with $\sigma_i$ being the Pauli matrices in the real spin space. The material parameters are listed in Appendix \ref{appxB}. In this case, the field operator associated with the Hamiltonian is given by $\hat{\psi}^\dag(\textbf{k})=(\psi_{A\ua}^\dag, \psi_{A\da}^\dag,\psi_{B\ua}^\dag, \psi_{B\da}^\dag)$, where the valleys and spins are labeled by $AB$ and $\ua\da$, respectively.

We assume that the superconductor can be described by the BCS formalism. There are four different scenarios for the coupling of particles in BP: 
\begin{subequations}
\begin{eqnarray}
& i) \;\;\;\; &\Delta^{AB}_{\ua\da}\Big\langle\psi^\dag_{A\ua}\psi^\dag_{B\da}\Big\rangle+\text{h.c.}\; ,\label{g1}\\
& ii) \;\;\;\; &\Delta^{AB}_{\ua\ua}\Big\langle\psi^\dag_{A\ua}\psi^\dag_{B\ua}\Big\rangle+\text{h.c.}\; ,\label{g2}\\
& iii) \;\;\;\; &\Delta^{AA,BB}_{\ua\da}\Big\langle\psi^\dag_{A\ua,B\ua}\psi^\dag_{A\da,B\da}\Big\rangle+\text{h.c.}\; ,\label{g3}\\
& iv) \;\;\;\; &\Delta^{AA,BB}_{\ua\ua}\Big\langle\psi^\dag_{A\ua,B\ua}\psi^\dag_{A\ua,B\ua}\Big\rangle+\text{h.c.}\; .\label{g4}
\end{eqnarray}
\end{subequations}
Pairings of type $i$) describe intervalley couplings with opposite spins, type $ii$) has intervalley couplings with equal spins, type $iii$) has intravalley couplings with opposite spins, and type $iv$) has intravalley couplings with equal spins. In general, the intravalley couplings can have different coupling potentials and therefore differing amplitudes in each valley.
The observation of intrinsic superconductivity in BP has attracted much attention \cite{Guo,Livas,Shirotani,Jun-JieZhang,YanqingFeng,Q.Huang} as well as controversies \cite{R.Zhang,Yuan}. Here we assume that the superconductivity can be induced extrinsically in the BP by making use of the proximity effect. In this paper we focus on only the intervalley spin-singlet coupling, i.e., $\Delta^{AB}_{\ua\ua}=\Delta^{AA,BB}_{\ua\da}=\Delta^{AA,BB}_{\ua\ua}=0$, and $\Delta^{AB}_{\ua\da}\neq 0$, which is most likely the energetically favored pairing in experiment \cite{beenakker}.

\begin{figure*}[t]
\includegraphics[width=18.0cm,height=5.30cm]{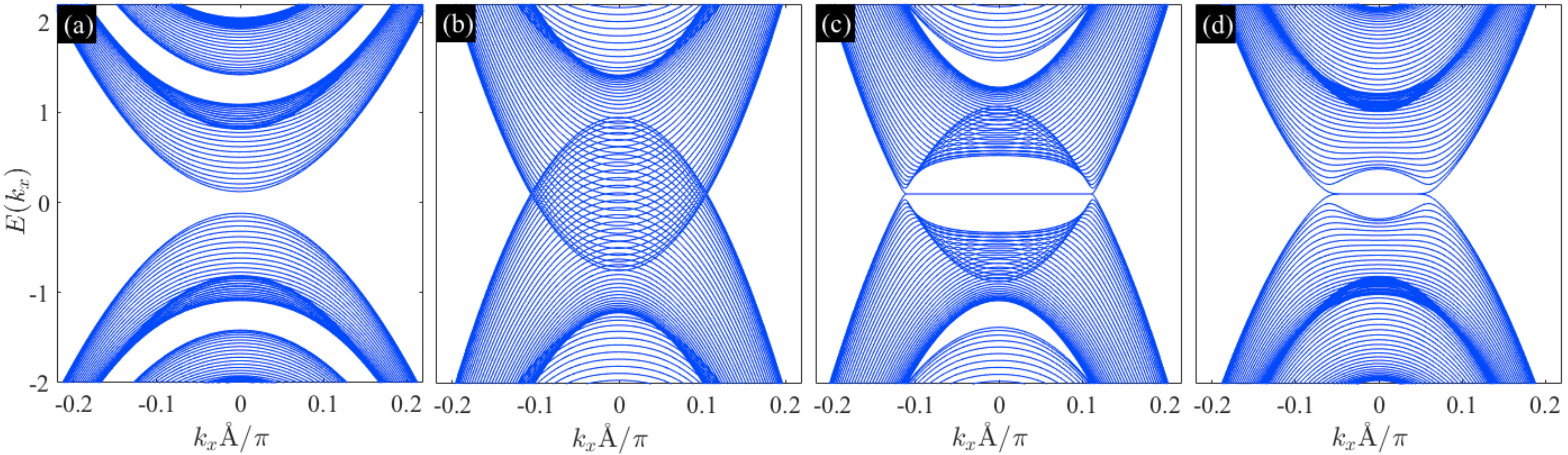}
\caption{\label{fig2} (Color online).
Band structure of a phosphorene monolayer, finite sized in the $y$ direction, as a function of momentum in the $x$ direction, i.e., $k_x$. (a) Strains $\varepsilon_{xx}, \varepsilon_{yy}$ are set to zero, and superconducting gap $|\Delta|\neq 0$. (b)  $\varepsilon_{xx}=-0.2$, $\varepsilon_{yy}=0$, and $|\Delta|=0$. (c) $\varepsilon_{xx}=-0.2$, $\varepsilon_{yy}=0$, and $|\Delta|\neq 0$. (d) $\varepsilon_{xx}=0$, $\varepsilon_{yy}=-0.2$, and $|\Delta|\neq 0$. }
\end{figure*}

We first discuss and analyze the band structure of BP with a superconducting gap. To this end, we use the Nambu space and write the Hamiltonian as
\begin{equation}\label{bcs}
{\cal H}(\textbf{k})=\left( \begin{array}{cc}
H(\textbf{k})-\mu & \hat{\Delta}\\
\hat{\Delta}^\dag & -H^{\text{T}}(-\textbf{k})+\mu
\end{array}\right),
\end{equation}
in which $\mu$ is the Fermi energy and we have suppressed the indices $A,B$ and $\ua\da$ so that $\Delta_{\ua\da}^{AB}$ $\rightarrow$ $\Delta$. The associated field operator can now be expressed by $\hat{\psi}_\text{BCS}^\dag(\textbf{k})=[\hat{\psi}^\dag(\textbf{k}),\hat{\psi}(\textbf{-k})]$. In the presence of intervalley singlet coupling, the dispersion relations of particle and hole branches can be obtained by diagonalizing the BCS Hamiltonian (\ref{bcs}): 
\begin{align}
\begin{split}
 &E_{e,h}(k_x,k_y)=\pm\Big( {\cal F}\pm 2\sqrt{\Xi}\Big)^{\frac{1}{2}},\\  &{\cal F}=|\Delta|^2+\Lambda^2+\Omega^2+\chi_y^2k_y^2, \\&\Xi=\Lambda^2(|\Delta|^2+\Omega^2)+\Omega^2\chi_y^2k_y^2,
 \end{split}
\end{align}
in which $\Omega=u_0+\alpha_i\varepsilon_{ii} +(\eta_j+\beta_{ij}\varepsilon_{ii})k_j^2+\mu$ and $\Lambda=\delta_0+\mu_i\varepsilon_{ii} +(\gamma_j+\nu_{ij}\varepsilon_{ii})k_j^2$.
The normal strain components $\varepsilon_{xx}$ and $\varepsilon_{yy}$ can take positive or negative values corresponding to expansion or compression,
respectively. Through the dispersion relation, one can examine where the band gap closes due to the strain, and the presence of $|\Delta|$ reopens a gap. However, to reveal the Majorana zero modes one needs to consider a system of finite size as in Fig. \ref{fig1}(a). The strain is assumed to be in the plane of BP. The Majorana fermions appear near the edges of the device. Therefore, by placing a scanning tunnel microscope (STM) tip right in the vicinity of the superconductor edge and BP, the Majorana zero mode shows up as a peak at zero bias. We also note that other methods such as point contact spectroscopy might be able to uncover the signatures of Majorana fermions through conductance data analyses \cite{Chang,Kjaergaard,Nichele,Deng,Mourik}.

The band parameters of BP under strain less than $25\%$ are obtained from density functional theory and symmetry computations \cite{Rudenko,Voon1,Voon2,Pereira,wei}, and are presented in Appendix \ref{appxB}. The superconducting gap for conventional bulk materials such as {\rm Al} is of the order of $\sim 0.2$ meV, and it can change when it is induced in a nonsuperconducting material by means of the proximity effect. Throughout our calculations, we set a nonzero superconducting gap $|\Delta|$ and present energies in units of $|\Delta|$. In order to retain simplicity in our analyses, we assume that the superconductivity in BP is (1) uniform and (2) finite sized in the $y$ direction with a representative relaxed length of $d_S\approx 50$~nm [see Fig. \ref{fig1}(a)]. We note that many-body effects in the presence of strain can change the electron-phonon interaction parameter and the screening of the electron-electron interaction potential. However, a detailed description of these fundamental phenomena, giving rise to superconductivity, goes beyond the scope of our paper, and therefore, we discard the effect of strain on the superconducting couplings. Nevertheless, as long as the strain does not change the type of pairing considered above, the findings in this paper are unaffected.

Throughout the main text, we set the chemical potential to zero, $\mu=0$; Appendixes \ref{appxB} and \ref{appxC} discuss the effects due to a finite $\mu$. By exact diagonalization of Eq. (\ref{bcs}) together with open boundaries at the edges, we find the band structure for the device shown in Fig. \ref{fig1}(a). In Fig. \ref{fig2}(a) the band structure is shown in the absence of strain $\varepsilon_{xx}=\varepsilon_{yy}=0$ and a nonzero superconducting gap $|\Delta|\neq 0$. As expected, the band structure possesses a direct gap (this happens also in the absence of $|\Delta|$, not shown). When the strain is large enough, we set a representative value $\varepsilon_{xx}=-0.2$ (equivalent to $20\%$ compression) and $|\Delta|=0$; the gap in the band structure of BP closes, and the top and bottom bands hybridize, as seen in Fig. \ref{fig2}(b). In Fig. \ref{fig2}(c), we switch the superconductivity on when the strain is nonzero ($\varepsilon_{xx}=-0.2$ and the gap of BP is closed). We see that a dispersionless flat band appears at zero energy and that it extends between the two created Dirac nodes at positive and negative $k_x$. The density of states related to this specific band demonstrates edge localized states in the finite-size superconductive BP and zero elsewhere (see Appendix \ref{appxB}). The zero energy mode inside the superconducting gap is called a Majorana flat band. The occurrence of this feature is not specific to the direction of the strain. In order to illustrate this, in Fig. \ref{fig2}(d) we set $\varepsilon_{xx}=0$ and $\varepsilon_{yy}=-0.2$ in the presence of superconductivity. The Majorana flat band appears in this case as well, except now its extension along $k_x$ is shortened. If the band parameters given in Appendix \ref{appxB} remain unchanged for very high strains, Weyl nodes with connecting Majorana flat bands, as well as single lines at finite energies, may appear (not shown). An experimentally more easily achievable scenario is to modify the Fermi level $\mu$; then flat bands appear at moderate strains $\varepsilon_{ii}\lesssim 25\%$ (Appendix \ref{appxB}, Fig. \ref{appfig1}). We emphasize that our conclusions in this paper remain the same for tensile stress lower than 20$\%$, as long as it keeps the band gap of BP closed. 

\begin{figure*}
  \centering
\includegraphics[ width=15.80cm,height=5.60cm]{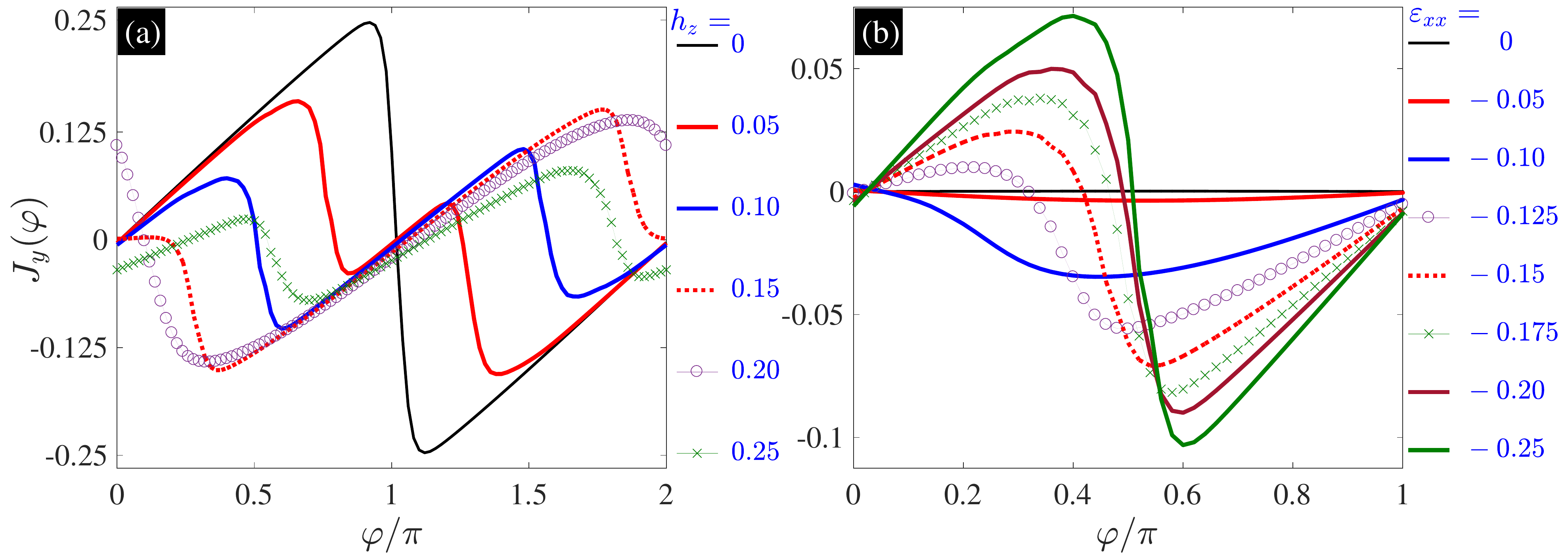}
\caption{ \label{fig3}(Color online).
Supercurrent density $J_y(\varphi)$ as a function of superconducting phase difference $\varphi$. (a) Different values of magnetization strength are considered: $h_z=0, 0.05, 0.1, 0.15, 0.2$, and $0.25$ eV, where we have set $\varepsilon_{xx}=-0.25$ and $\varepsilon_{yy}=0$. (b) Magnetization strength is set to $h_z=0.1$ eV, and the current density is examined for differing values of $\varepsilon_{xx}=0, -0.05, -0.1, -0.125, -0.15, 0.175, -0.2$, and $-0.25$ where $\varepsilon_{yy}=0$.   }
\end{figure*}

The Fermi level can be tuned by a gate electrode or an {\it in situ} deposition of {\rm K} or {\rm Rb} atoms on BP \cite{kim1,kim2}. The gap-closing condition can be determined by the chiral symmetry transformation $U$ (made of particle-hole and time-reversal operators), so that
\begin{equation}
U{\cal H}(\textbf{k})U^\dag=\left(\begin{array}{cc}
0 & {\cal G}(\textbf{k})\\
{\cal G}^\dag(\textbf{k}) &0
\end{array}\right), \;\;\;\; U = \frac{1}{\sqrt{2}}(\tau_z+\tau_x)\sigma_x,
\end{equation}
where ${\cal G}(\textbf{k})$ $=$ $-$ $\Omega^2\sigma_0$ $-$ $\Lambda^2\sigma_x$ $-$ $(\chi_yk_y+i\Delta)\sigma_y$. Since det{${\cal H}(\textbf{k})$}$=$ $-$ det{${\cal G}(\textbf{k})$} det{${\cal G}^\dag(\textbf{k})$}, the solution to det{${\cal G}(\textbf{k})$}=$\Omega^2 - \Lambda^2-(\chi_yk_y+i\Delta)^2$=0 determines the gap-closing condition in the parameter space (see Appendix \ref{appxB}).

In order to study the supercurrent behavior in the device depicted in Fig. \ref{fig1}(b), we compute the current directly through its quantum definition: 
\begin{equation}
{\bf J}=\int \dr \Big\{ \hat{\psi}^\dag({\bm r}) \overrightarrow{{\cal H}}({\bm r})\hat{\psi}({\bm r}) - \hat{\psi}^\dag({\bm r}) \overleftarrow{{\cal H}}^\dag({\bm r})\hat{\psi}({\bm r})\Big\},
\end{equation}
where ${\cal H}({\bm r})$ is given by Eq. (\ref{bcs}), using ${\bf k} \rightarrow - i {\bm \nabla}$ [see Eq. (\ref{haml_supp}) in Appendix \ref{appxB}], and the arrows show the direction of derivatives. The current density is in units of the elementary charge and superconducting gap. Thus, to obtain the total charge current in a junction of width $W_F$, one must multiply the current density by $2e|\Delta| W_F/h$. The application of current conservation law results in the continuity of wave functions at the interfaces of a junction with left ($l$) and right (r) segments $\hat{\psi}_l=\hat{\psi}_r$ together with $\partial_\textbf{k} {\cal H}_l(\textbf{k})\hat{\psi}_l=\partial_\textbf{k} {\cal H}_r(\textbf{k})\hat{\psi}_r$. The wave functions associated with the electron and hole branches are given in Appendix \ref{appxA}. Without any assumptions regarding the location of the Fermi level the resulting expressions are very complicated but, nevertheless, readily evaluated numerically. The length of the relaxed ferromagnetic BP in Fig. \ref{fig1}(b) is set to $d_F\approx 7.5$~nm, which is a representative value in the tunneling regime. Figure \ref{fig3} shows the charge current density as a function of phase difference ($\varphi=\theta_l-\theta_r$) between the macroscopic phases of the left and right superconductors' wave functions ($\theta_{l,r}$) shown in Fig. \ref{fig1}(b). We consider an applied strain throughout the junction so that $\varepsilon_{xx}=-0.2$ and $\varepsilon_{yy}=\mu=0$ with a uniform magnetization directed along the $z$ axis, $h_z$. In Fig. \ref{fig3}(a)
we show the current vs. $\varphi$ for various magnetic fields. When the exchange field is zero, the supercurrent shows the conventional nonsinusoidal current phase relation in the ballistic limit. When the exchange field is nonzero, higher harmonics appear. By increasing the exchange field, the second harmonic dominates, while the amplitude of the supercurrent decreases until the $0$ to $\pi$ transition is complete. At the same time, the supercurrent experiences a phase shift $\varphi_0$, which depends on $h_z$ and is pronounced at large enough values of $h_z$. Note that the exchange field is a Cooper pair breaking factor. Therefore, by increasing the exchange field, while $\varphi_0$ is enhanced, the amplitude of the total current is exponentially suppressed. The $\varphi_0$ Josephson state was also discussed in other contexts \cite{zyuzin,bobkova,alidoustphi0,Krive,Shapiro} and recently realized in experiment \cite{Szombati,roso,herve}. In Fig. \ref{fig3}(b) we set $h_z=0.1$ eV, $\varepsilon_{yy}=\mu=0$, and vary the strain in the $x$ direction. When $\varepsilon_{xx}$ is zero, the supercurrent is vanishingly small. This can be understood by noting that in this parameter regime where $\varepsilon_{ii}=\mu=0$ the BP is fully gapped. By increasing $\varepsilon_{xx}$, the gap starts to close, and hence, the supercurrent begins to increase. When $\varepsilon_{xx}=-0.1$, the supercurrent is negative almost throughout the phase difference interval $\varphi=[0,\pi]$. By increasing the compressive strain to $\varepsilon_{xx}=-0.125$, the higher harmonics emerge gradually, and the supercurrent changes sign. A further increase in the magnitude of $\varepsilon_{xx}$ makes the supercurrent reversal more pronounced. Therefore, the external strain can control the supercurrent reversal, the appearance of higher harmonics, and the phase shift $\varphi_0$. For further investigations of the influence of the strain tensor in the presence of a finite chemical potential on the supercurrent, see Appendix \ref{appxC}.

\section{conclusions}

In conclusion, we have shown that a black phosphorus monolayer subject to external strain hosts a variety of phenomena in the presence of superconductivity. Specifically, our results reveal that a finite-size superconductive BP exhibits strain-driven Majorana zero energy edge modes. Manipulating the applied strain and chemical potential, one is able to create Dirac and Weyl nodes with connecting dispersionless zero energy flat bands. Thus, superconductive BP offers an experimental platform for studying Dirac and Weyl superconductivity. We also study the behavior of supercurrent in a ferromagnetic BP Josephson junction. We find that the magnetization induces higher harmonics, i.e., $\sin n\varphi$, $n=\pm 2, \pm 3, ...$ in the current phase relation and causes nonzero current at $0$ to $\pi$ crossovers. The strain tunes the higher harmonics and induces supercurrent reversals. In addition, the supercurrent experiences a phase shift $\varphi_0$, causing a spontaneous supercurrent at zero superconducting phase difference. Our findings demonstrate that black phosphorous offers a topological strain-effect transistor which is an excellent framework for studying a number of exotic transport phenomena.

\begin{acknowledgments}
M.A. is supported by Iran's National Elites Foundation (INEF). Center for Nanostructured Graphene is supported by the Danish National Research Foundation (Project No. DNRF103).
\end{acknowledgments}

\section{appendix}

\appendix

\begin{figure*}[t]
\includegraphics[width=18.0cm,height=5.30cm]{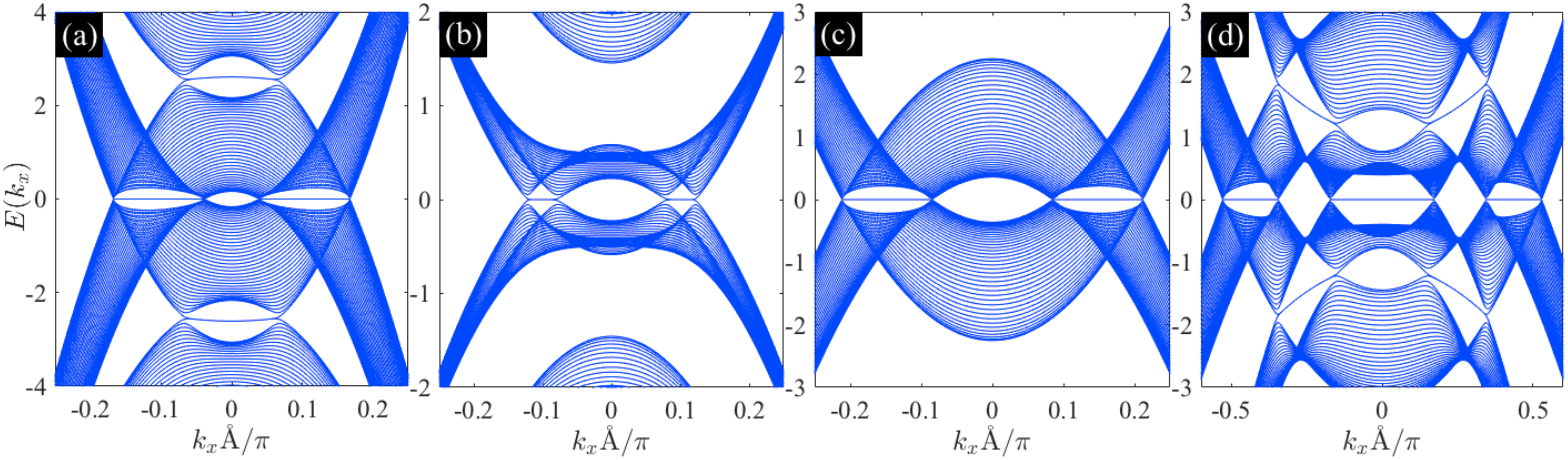}
\includegraphics[width=18.0cm,height=4.50cm]{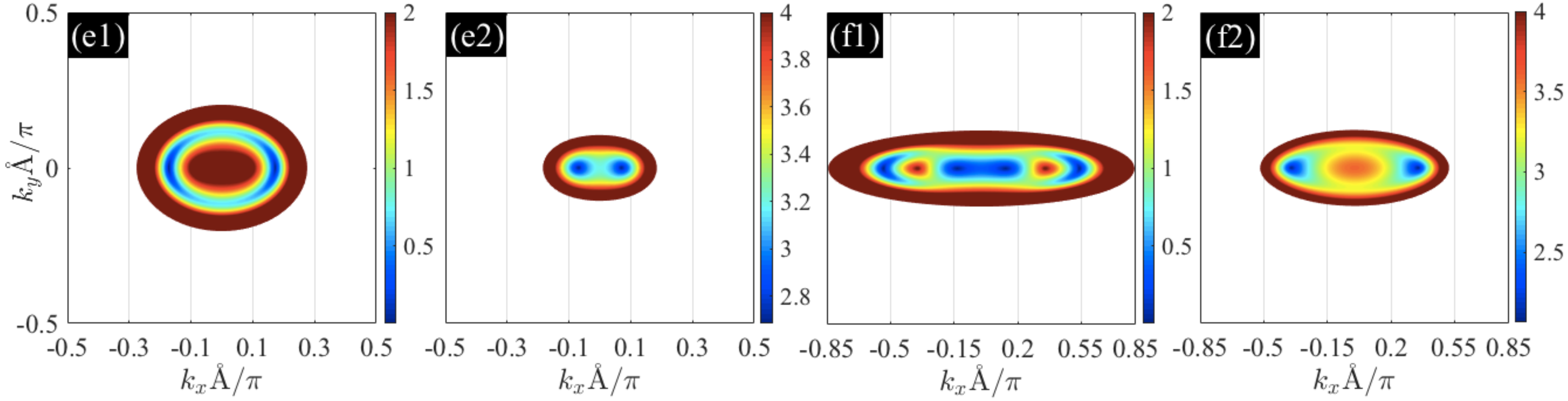}
\includegraphics[ width=12.80cm,height=5.60cm]{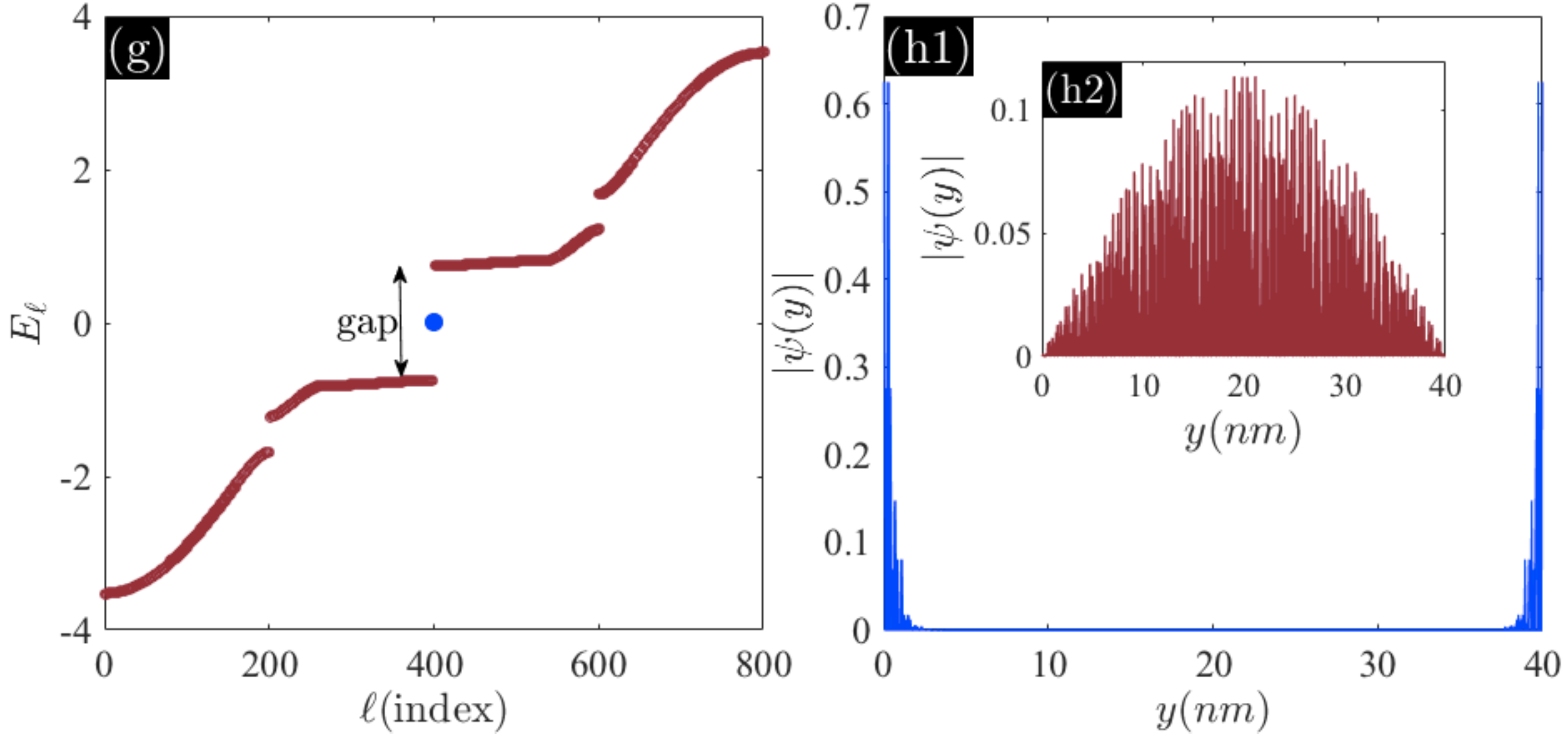}
\caption{\label{appfig1} (Color online).
(a)-(d): Band structure of finite-size black phosphorus in the $y$ direction as a function of momentum in the $x$ direction, i.e., $k_x$. The parameters are set equal to those of Fig. \ref{fig2}, except now the chemical potential possesses a nonzero value, $\mu=1.5$ eV. (a) $\varepsilon_{xx}=-0.25$ and $\varepsilon_{yy}=-0.25$. (b) $\varepsilon_{xx}=+0.25$ and $\varepsilon_{y}=-0.25$. (c) $\varepsilon_{xx}=-0.25$ and $\varepsilon_{yy}=+0.25$. (d) $\varepsilon_{xx}=+0.25$ and $\varepsilon_{yy}=+0.25$. (e1) and (e2): Band structure of superconductive BP with parameters set identical to those in (a), except now no boundaries are applied to the system. (f1) and (f2): Band structure of infinite superconductive BP without any boundaries with parameters set identical to those in (d). In (g), (h1), and (h2)  we set $|\Delta|\neq 0$, $d_S=40$~nm, $\varepsilon_{xx}=-0.2$, and $\varepsilon_{yy}=k_x=0$. (g) Eigenenergies $E_{\ell}$ vs the energy index ($\ell$). (h1) and (h2): The absolute wave function $|\psi(y)|$ vs the location $y$(nm)$\in [0,d_S]$ inside the system, corresponding to different eigenenergies presented in (g).}
\end{figure*}


In the appendixes, we first present the eigenvalues and associated wave functions in the normal and superconductive BP (Appendix A). Next, we give band parameters and extend our study in the main text to further analyze the band structure of the superconductive BP (Appendix B). Finally, we study the influence of $\varepsilon_{xx}$, $\varepsilon_{yy}$, and $\mu$ on the supercurrent behavior (Appendix C).

\section{Wave functions}\label{appxA}

By directly diagonalizing Eq. (\ref{bcs}) in the main text we find the following wave functions for the electrons $\hat{\psi}_e$ and holes $\hat{\psi}_h$ propagating along the $y$ direction in the plane of BP in the presence of singlet superconductivity with a gap of $\Delta$:
\begin{equation}
\begin{split}
&\hat{\psi}_h=\Big(f_1,f_2,f_3,1\Big)^\text{T}e^{ik_yy},\\
&f_1=\frac{\chi_y k_y (\zeta-\Omega (\Omega+\varepsilon_-))+i\Lambda (\Delta^2-\zeta+\Omega (\Omega+\varepsilon_-))}{\Delta (\chi_yk_y\Omega-i\Lambda \varepsilon_-)},\\
&f_2=i\frac{\chi_y^2 k_y^2\Omega+\Lambda^2\Omega-\zeta(\Omega+\varepsilon_-)}{\Delta (\chi_yk_y\Omega-i\Lambda \varepsilon_-)},\\
&f_3=-\frac{i\chi_y k_y \Lambda-\Lambda^2+\zeta}{i \chi_y k_y\Omega +\Lambda\varepsilon_-},
\end{split}
\end{equation}
and
\begin{equation}
\begin{split}
&\hat{\psi}_e=\Big(g_1,g_2,g_3,1\Big)^\text{T}e^{ik_yy},\\
&g_1=\frac{-\chi_y k_y (\zeta+\Omega (\Omega+\varepsilon_+))+i\Lambda (\Delta^2+\zeta+\Omega (\Omega+\varepsilon_+))}{\Delta (\chi_yk_y\Omega+i\Lambda \varepsilon_+)},\\
&g_2=i\frac{\chi_y^2 k_y^2\Omega+\Lambda^2\Omega+\zeta(\Omega+\varepsilon_+)}{\Delta (\chi_y k_y\Omega+i\Lambda \varepsilon_+)},\\
&g_3=i\frac{i\chi_y k_y \Lambda+\Lambda^2+\zeta}{ \chi_y k_y\Omega -i\Lambda\varepsilon_+},
\end{split}
\end{equation}
with the associated eigenvalues
\begin{eqnarray}
\varepsilon_{\pm}=\sqrt{\chi_y^2 k_y^2+\Lambda^2+\Omega^2+\Delta^2\pm 2\zeta},
\end{eqnarray}
in which we have defined
\begin{eqnarray}
\zeta=\sqrt{\chi_y^2k_y^2\Omega^2+\Lambda^2(\Delta^2+\Omega^2)},
\end{eqnarray}
and T stands for the matrix transpose operation. The wave functions in the nonsuperconducting region for the propagating electrons and holes become simpler in the absence of $\Delta$:
\begin{eqnarray}
&\hat{\psi}_e=\Big(\frac{-i\chi_y k_y + \Lambda}{\sqrt{\chi_y^2 k_y^2 + \Lambda^2}},1,0,0\Big)^\text{T}e^{-ik_yy},\\
&\hat{\psi}_h=\Big(0,0,\frac{i\chi_y k_y - \Lambda}{\sqrt{\chi_y^2 k_y^2 + \Lambda^2}},1\Big)^\text{T}e^{-ik_yy},
\end{eqnarray}
and the associated eigenvalues are given by
\begin{eqnarray}
&\varepsilon_e=\sqrt{\chi_y^2 k_y^2 + \Lambda^2} + \Omega,\\
&\varepsilon_h=\sqrt{\chi_y^2 k_y^2 + \Lambda^2} - \Omega,
\end{eqnarray}
where we have defined
\begin{eqnarray}
&\Omega=u_0+\alpha_i\varepsilon_{ii} +(\eta_j+\beta_{ij}\varepsilon_{ii})k_j^2,\\
&\Lambda=\delta_0+\mu_i\varepsilon_{ii} +(\gamma_j+\nu_{ij}\varepsilon_{ii})k_j^2.
\end{eqnarray}

\begin {table}[t]
\caption {Band parameters of the phosphorene monolayer subject to externally applied strain \cite{Voon1,Voon2}. } \label{table}
\begin{center}
\begin{tabular}{c*{4}{c}c}
\hline
\hline
$u_0$(eV) & $\delta_0$(eV) & $\alpha_x$(eV) & $\alpha_y$(eV) & $\mu_x$(eV)    \\
-0.42  & +0.76  & +3.15  & -0.58  & +2.65     \\
\hline
 $\mu_y$(eV) & $\eta_x$(eV$ \textup{\AA}^2$)  & $\eta_y$(eV$ \textup{\AA}^2$)  & $\gamma_x$(eV$ \textup{\AA}^2$)  & $\gamma_y$(eV$ \textup{\AA}^2$) \\
 +2.16  &  +0.58  & +1.01 & +3.93 & + 3.83  \\
\hline
$\beta_{xx}$(eV$ \textup{\AA}^2$) & $\beta_{yx}$(eV$ \textup{\AA}^2$) & $\beta_{xy}$(eV$ \textup{\AA}^2$) & $\beta_{yy}$(eV$ \textup{\AA}^2$)  \\
-3.48 & -0.57 & +0.80 & +2.39\\
\hline
$\nu_{xx}$(eV$ \textup{\AA}^2$)  & $\nu_{yx}$(eV$ \textup{\AA}^2$) & $\nu_{xy}$(eV$ \textup{\AA}^2$) & $\nu_{yy}$(eV$ \textup{\AA}^2$) & $\chi_y$(eV$ \textup{\AA}$)  \\
-10.90 & -11.33 & -41.40 & -14.80 & +5.25\\
\hline
\hline
\end{tabular}
\end{center}
\end{table}

\section{Band structure analyses}\label{appxB}

\begin{figure*}[t]
\includegraphics[width=18.0cm,height=5.70cm]{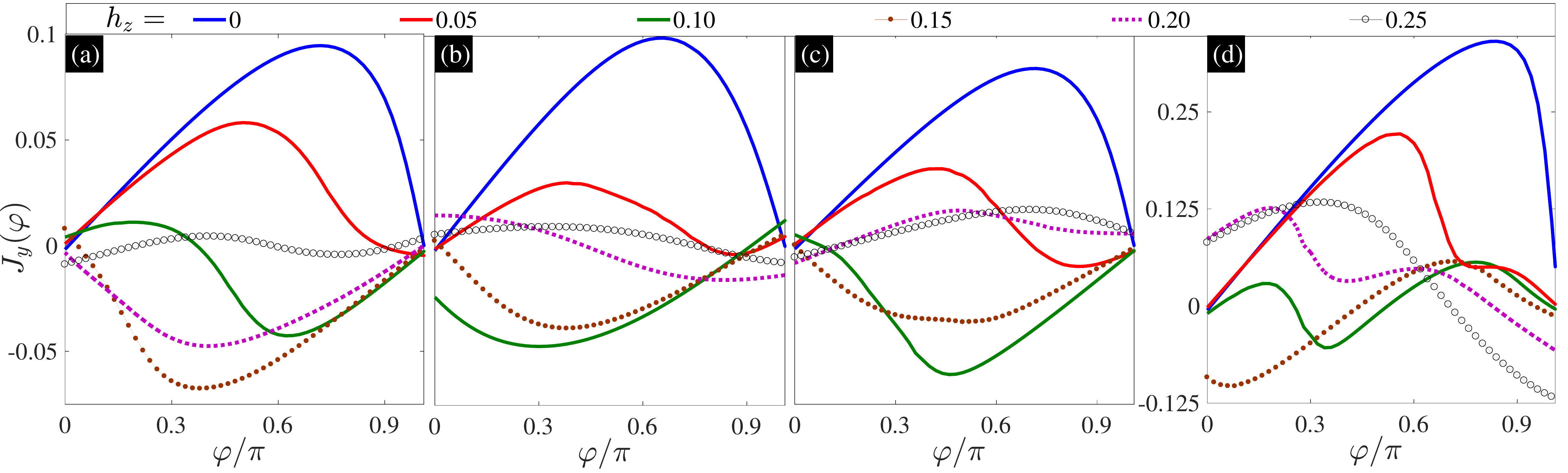}
\caption{\label{appfig2} (Color online).
Supercurrent density $J_y(\varphi)$ in the $y$ direction for various values of magnetization strength $h_z=0, 0.05, 0.1, 0.15, 0.2$ eV, and $0.25$ eV, where we have fixed the chemical potential at $\mu=1.5$ eV. (a) $\varepsilon_{xx}=-0.25$ and $\varepsilon_{yy}=-0.25$. (b) $\varepsilon_{xx}=+0.25$ and $\varepsilon_{yy}=-0.25$. (c) $\varepsilon_{xx}=-0.25$ and $\varepsilon_{yy}=+0.25$. (d) $\varepsilon_{xx}=+0.25$ and $\varepsilon_{yy}=+0.25$.}
\end{figure*}

To gain more insight into the influences of strain on Dirac and Weyl nodes as well as the dispersionless flat bands, we plot the band structure $E(k_x)$ of finite-size superconductive BP in Fig. \ref{appfig1}, where the confining edges are located in the $y$ direction so that particle momentum in the $x$ direction $k_x$ is a good quantum number. The Hamiltonian in real space takes the following form:
\begin{eqnarray}\label{haml_supp}
H({\bf r})&=& \big [u_0+\alpha_i\varepsilon_{ii} -(\eta_j+\beta_{ij}\varepsilon_{ii})\nabla_j^2\big]\tau_0 +\nonumber\\&+&\big[\delta_0+\mu_i\varepsilon_{ii} -(\gamma_j+\nu_{ij}\varepsilon_{ii})\nabla_j^2\big ]\tau_x +i\chi_y \nabla_y\tau_y.
\end{eqnarray}
Also, we obtained the band parameters of BP through density functional theory and symmetry computations. These band parameters are given in Table \ref{table}. The rest of the parameters are identical to those of Fig. \ref{fig2} in the main text, except now we set a representative finite chemical potential $\mu=1.5$ eV. Figure \ref{appfig1}(a) exhibits $E(k_x)$, where $\varepsilon_{xx}=-0.25$ and $\varepsilon_{yy}=-0.25$. We see that the Dirac points in Fig. 2(c) or 2(d) split into two Weyl nodes around $k_x=0$. The Weyl nodes on each side are connected by a flat band, and  its associated wave function is localized at the edges in the $y$ direction (see below). Also, two Dirac points at finite energies appear that are connected by a single band. This picture remains intact in cases where $\varepsilon_{xx}=+0.25$, $\varepsilon_{yy}=-0.25$ and $\varepsilon_{xx}=-0.25$, $\varepsilon_{yy}=+0.25$, as shown in Figs. \ref{appfig1}(b) and \ref{appfig1}(c). As seen in Fig. \ref{appfig1}(d), the case $\varepsilon_{xx}=+0.25$, $\varepsilon_{yy}=+0.25$ exhibits six Weyl nodes at zero energy distributed equally around $k_x=0$. The two inner nodes are connected by a flat band, while the outer two nodes 
located at positive and negative values of $k_x$ are separately connected by two flat bands. Also, we see that the Dirac nodes at finite energies are now split
 into two Weyl nodes shifted in energy and yet connected by single bands. Interestingly, the outer Weyl nodes are tilted to some extent and are suggestive of creating type-II Weyl nodes by proper combinations of strain and finite chemical potential \cite{Li,Hou,Xiao,Alidoust,Sun,Soluyanov_app,Ruan_app,Marra1,wylsuperc_exp1}. Figures \ref{appfig1}(e1) and \ref{appfig1}(e2) exhibit the low energy band structure as a function of $k_x$ and $k_y$ where no boundaries are imposed and $k_x$ and $k_y$ are good quantum numbers. The band parameter values are set identical to those of Fig. \ref{appfig1}(a). A similar study is exhibited in Figs. \ref{appfig1}(f1) and \ref{appfig1}(f2) with parameters corresponding to those of Fig. \ref{appfig1}(d). As seen, the gap closes at $k_y=0$ and specific values of $k_x$. To obtain exact values of $k_x$ where the gap closes, we use det{${\cal G}(\textbf{k})$}=$\Omega^2 - \Lambda^2-(\chi_yk_y+i\Delta)^2$=0 given in the main text and set $k_y=0$. After 
straightforward calculations we find:
\begin{widetext}
\begin{equation}
k_x=\pm \sqrt{\frac{\Omega_0\Omega_x-\Lambda_0\Lambda_x\pm\sqrt{ \Lambda_x^2(\Omega_0^2+\Delta^2) -2\Lambda_0\Lambda_x\Omega_0\Omega_x +\Omega_x^2(\Lambda_0^2-\Delta^2)}}{\Lambda_x^2-\Omega_x^2 }},\\
\end{equation}
\end{widetext}
where we have defined $\Omega_0= u_0 +\alpha_i\varepsilon_{ii} -\mu$, $\Omega_x = \eta_x+\beta_{ix}\varepsilon_{ii}$, $\Lambda_0 =\delta_0+\mu_i\varepsilon_{ii}$, $\Lambda_x= \gamma_x+\nu_{ix}\varepsilon_{ii}$.

To illustrate the edge localized zero energy states, we have plotted the eigenenergies $E_{\ell}$ vs the energy index ($\ell$) in Fig. \ref{appfig1}(g). The set up is shown in Fig. 1(a) in the main text. We have set $|\Delta|\neq 0$, $d_S=40$~nm, $\varepsilon_{xx}=-0.2$, and $\varepsilon_{yy}=k_x=0$, where the edges are located in the $y$ direction at $y=0, 40$~nm. The corresponding band structure is shown in Fig. \ref{fig2}(c) of the main text. As seen, in Fig. \ref{appfig1}(g), there is a gap in the eigenenergies at $\ell=400$ with ingap states at zero energy indicated by the blue circle. In other words, this zero energy state is well separated from the other states displayed by the red circles, indexed from $0$ to $800$. To show the difference between the zero energy state and other states, we plot the corresponding wave functions $|\psi(y)|$, which is proportional to the local density of states, in Figs. \ref{appfig1}(h1) and \ref{appfig1}(h2), respectively. We see that the state at zero energy is highly localized at the edges $y=0,40$~nm and rapidly decays to zero when moving away from the edges. The other states propagate throughout the system and vanish at the edges $y=0,40$~nm as seen in Fig. \ref{appfig1}(h2) for a representative state.

\section{Supercurrent}\label{appxC}
Finally, we study the influence of strain on the supercurrent in a BP-based ferromagnetic Josephson junction with a finite chemical potential. Figure \ref{appfig2} shows the supercurrent density as a function of superconducting phase difference $\varphi$ where the chemical potential is fixed at $\mu=1.5$ eV and the exchange field varies ($h_z=0, 0.05, 0.1, 0.15, 0.2$, and $0.25$ eV). In Fig. \ref{appfig2}(a) we set $\varepsilon_{xx}=-0.25$, $\varepsilon_{yy}=-0.25$. 
The supercurrent has a small deviation from zero at $\varphi=0$ for $h_z=0.05$~eV. When the magnetization is equal to $0.1$ eV, we see that the supercurrent reversal is apparent now due to the stronger contribution of higher harmonics and the nonzero supercurrent at $\varphi=0$ grows as well. At $h_z=0.15$ eV, the contribution of higher harmonics is again less than the first harmonic, except now the current has overall changed its sign compared to the case where $h_z=0$. By increasing the magnetization strength to $h_z=0.2$ eV, the finite supercurrent at zero phase difference is highly suppressed in addition to the suppression of higher harmonics. By further increasing $h_z$ to $0.25$ eV, these two phenomena show up again, except now the amplitude of the supercurrent is highly suppressed by the magnetization itself. In Fig. \ref{appfig2}(b) we set $\varepsilon_{xx}=+0.25$, $\varepsilon_{yy}=-0.25$. We see that the appearance of higher harmonics and the $\varphi_0$ phase shift in the presence of weak magnetization is more 
pronounced now. Specifically, the supercurrent experiences a fairly large nonzero value at zero phase difference $\varphi=0$ if $h_z = 0.10$ or $0.20$~eV. 
Figure \ref{appfig2}(c) shows that when $\varepsilon_{xx}=-0.25$, $\varepsilon_{yy}=+0.25$, the supercurrent can change direction compared to the previous case in Fig. \ref{appfig2}(b). Additionally, the $\varphi_0$ phase shift is now suppressed even more than Fig. \ref{appfig2}(a). Similar to the band structure shown in Fig. \ref{appfig1}(d), the supercurrent shows richer features when $\varepsilon_{xx}=+0.25$, $\varepsilon_{yy}=+0.25$. The corresponding supercurrent is plotted in Fig. \ref{appfig2}(d). We see that for nonzero values of magnetization strength, the higher-harmonic contributions to the supercurrent dominate. Additionally, the spontaneous supercurrent at zero phase difference for nonzero magnetizations is stronger than in the previous cases illustrated in Figs. \ref{appfig2}(a)-\ref{appfig2}(c).


\begin{thebibliography}{000}

\bibitem{kane_rmp} M. Z. Hasan and C. L. Kane, {\it Colloquium: topological insulators}, \href{https://journals.aps.org/rmp/abstract/10.1103/RevModPhys.82.3045}{Rev. Mod. Phys. {\bf 82}, 3045 (2010)}.

\bibitem{zhang_rmp} X.-L. Qi and S.-C. Zhang, {\it Topological insulators and superconductors}, \href{https://journals.aps.org/rmp/abstract/10.1103/RevModPhys.83.1057}{Rev. Mod. Phys. {\bf 83}, 1057 (2011)}.

\bibitem{Beenakker2013ARCM}
C. W. J. Beenakker, {\it Search for Majorana fermions in superconductors}, \href{http://www.annualreviews.org/doi/10.1146/annurev-conmatphys-030212-184337}{ Ann. Rev. Cond. Matt. \textbf{4}, 113 (2013)}.

\bibitem{Nayak2008RMP}
C. Nayak, S. H. Simon, A. Stern, M. Freedman, and S. Das Sarma, {\it Non-Abelian anyons and topological quantum computation}, \href{http://journals.aps.org/rmp/abstract/10.1103/RevModPhys.80.1083}{Rev. Mod. Phys. \textbf{80}, 1083 (2008)}.

\bibitem{Maeno} Y. Maeno, T.M. Rice. M. Sigrist, {\it The intriguing superconductivity of strontium ruthenate}, \href{https://repository.kulib.kyoto-u.ac.jp/dspace/bitstream/2433/49957/1/PTO000042.pdf}{Physics Today \textbf{54}, 42
(2001)}.


\bibitem{Mackenzie} A.P. Mackenzie and Y. Maeno, {\it The superconductivity of  ${\rm Sr_2RuO_4} $ and the physics of spin-triplet pairing}, \href{https://journals.aps.org/rmp/abstract/10.1103/RevModPhys.75.657}{Rev. Mod. Phys. \textbf{75}, 657
(2003)}.

\bibitem{Liu} Y.-C. Liu, F.-C. Zhang, T. M. Rice and Q.-H. Wang, {\it Theory of the evolution of superconductivity in ${\rm Sr_2 RuO_4}$ under anisotropic strain}, \href{https://www.nature.com/articles/s41535-017-0014-y}{npj Quantum Mater. \textbf{2}, Article number: 12 (2017)}.

\bibitem{Zhang1} J. Zhang, W. Huang, M. Sigrist, and D. Yao, {\it Interband interference effects at the edge of a multiband chiral-wave superconductor}, \href{https://journals.aps.org/prb/abstract/10.1103/PhysRevB.96.224504}{Phys. Rev. B \textbf{96}, 224504 (2017)}.

\bibitem{Gregoras} D. Gregoras, G. Papini, {\it Weyl-Dirac theory and superconductivity}, \href{https://link.springer.com/article/10.1007/BF02755094}{Nuov Cim B \textbf{63}, 487 (1981)}.

\bibitem{law}  W. He, B. T. Zhou, J. J. He, N. Yuan, T. Zhang, K. T. Law, {\it Magnetic Field Driven Nodal Topological Superconductivity in Monolayer Transition Metal Dichalcogenides}, \href{https://arxiv.org/abs/1604.02867}{arXiv:1604.02867}

\bibitem{OBrien} T. E. OBrien, C. W. J. Beenakker, and İ. Adagideli, {\it Superconductivity Provides Access to the Chiral Magnetic Effect of an Unpaired Weyl Cone}, \href{https://journals.aps.org/prl/abstract/10.1103/PhysRevLett.118.207701}{Phys. Rev. Lett. \textbf{118}, 207701 (2017)}.

\bibitem{Li} D. Li, B. Rosenstein, B. Ya. Shapiro, and I. Shapiro, {\it Effect of the type I to type II Weyl semimetal topological transition on superconductivity}, \href{https://journals.aps.org/prb/abstract/10.1103/PhysRevB.95.094513}{Phys. Rev. B \textbf{95}, 094513 (2017)}.

\bibitem{Hou} Z. Hou and Q. Sun, {\it Double Andreev Reflections in Type-II Weyl Semimetal-Superconductor Junctions}, \href{https://journals.aps.org/prb/abstract/10.1103/PhysRevB.96.155305}{Phys. Rev. B \textbf{96}, 155305 (2017)}.

\bibitem{Alidoust} M. Alidoust, K. Halterman, and A. A. Zyuzin, {\it Superconductivity in type-II Weyl semimetals}, \href{https://journals.aps.org/prb/abstract/10.1103/PhysRevB.95.155124}{Phys. Rev. B \textbf{95}, 155124 (2017)}.


\bibitem{Sun} H.-H. Sun, K.-W. Zhang, L.-H. Hu, C. Li, G.-Y. Wang,
H.-Y. Ma, Z.-A. Xu, C.-L. Gao, D. Guan, Y.-Y. Li, C. Liu, D. Qian, Y. Zhou, L. Fu, S.-C. Li, F.-C. Zhang, J.-F. Jia, {\it Majorana zero mode detected with spin selective Andreev reflection in the vortex of a topological superconductor}, \href{https://journals.aps.org/prl/abstract/10.1103/PhysRevLett.116.257003}{Phys. Rev. Lett. \textbf{116}, 257003 (2016)}.

\bibitem{Das} S. Das, Amit, A. Sirohi, L. Yadav, S. Gayen, Y. Singh, and G. Sheet, {\it Conventional Superconductivity in Type-II Dirac Semimetal PdTe2}, \href{https://journals.aps.org/prb/abstract/10.1103/PhysRevB.97.014523}{Phys. Rev. B \textbf{97}, 014523 (2018)}.

\bibitem{Huang} B. Huang, X. Yang, N. Xu, and M. Gong, {\it Type-I and type-II topological nodal superconductors with s-wave interaction}, \href{https://journals.aps.org/prb/abstract/10.1103/PhysRevB.97.045142}{Phys. Rev. B \textbf{97}, 045142 (2018)}.

\bibitem{Izumida} W. Izumida, L. Milz, M. Marganska, and M. Grifoni, {\it Topology and zero energy edge states in carbon nanotubes with superconducting pairing
}, \href{https://journals.aps.org/prb/abstract/10.1103/PhysRevB.96.125414}{Phys. Rev. B \textbf{96}, 125414 (2017)}.

\bibitem{Zhang} P. Zhang, K. Yaji, T. Hashimoto, Y. Ota, T. Kondo, K. Okazaki, Z. Wang, J. Wen, G. D. Gu, H. Ding, S. Shin, {\it Observation of topological superconductivity on the surface of iron-based superconductor}, \href{https://arxiv.org/abs/1706.05163}{arXiv:1706.05163}.

\bibitem{Lv} Y. Lv, W. Wang, Y. Zhang, H. Ding, W. Li, L. Wang, K. He, C. Song, X. Ma, and Q. Xue, {\it Experimental Signature of Topological Superconductivity and Majorana Zero Modes on ${\rm \beta-Bi_2Pd}$ Thin Films}, \href{https://www.sciencedirect.com/science/article/pii/S2095927317302487}{Science Bulletin \textbf{62}, 852 (2017)}.

\bibitem{Bachmann} M. D. Bachmann, N. Nair, F. Flicker, R. Ilan, T. Meng, N. J. Ghimire, E. D. Bauer, F. Ronning, J. G. Analytis, P.  Moll, {\it Inducing superconductivity in Weyl semimetal microstructures by selective ion sputtering}, \href{http://advances.sciencemag.org/content/3/5/e1602983.abstract}{Sci. Adv. \textbf{3}, 5 (2017)}.

\bibitem{Xiao} R. C. Xiao, P. L. Gong, Q. S. Wu, W. J. Lu, M. J. Wei, J. Y. Li, H. Y. Lv, X. Luo, P. Tong, X. B. Zhu, and Y. P. Sun, {\it Experimental Realization of Type-II Dirac Fermions in ${\rm PdTe_2}$ Superconductor}, \href{https://journals.aps.org/prb/abstract/10.1103/PhysRevB.96.075101}{Phys. Rev. B \textbf{96}, 075101 (2017)}.

\bibitem{Oudah} M. Oudah, A. Ikeda, J. Hausmann, S. Yonezawa, T. Fukumoto, S. Kobayashi, M. Sato and Y. Maeno, {\it Superconductivity in the antiperovskite Dirac-metal oxide ${\rm Sr_{3-x}SnO}$}, \href{https://www.nature.com/articles/ncomms13617}{Nat. Comm. \textbf{7}, Article number: 13617 (2016)}.

\bibitem{He} L. He, Y. Jia, S. Zhang, X. Hong, C. Jin and S. Li, {\it Pressure-induced superconductivity in the three-dimensional topological Dirac semimetal Cd3As2}, \href{https://www.nature.com/articles/npjquantmats201614}{npj Quantum Mater. \textbf{1}, 16014 (2016)}.

\bibitem{volov1} G.E. Volovik, {\it The Universe in a Helium Droplet}, \href{}{Clarendon Press, Oxford (2003)}; G. E. Volovik, {\it Exotic Lifshitz transitions in topological materials}, \href{https://arxiv.org/abs/1701.06435}{	arXiv:1701.06435}; J. Nissinen and G. E. Volovik, {\it Type-III and IV interacting Weyl points}, \href{https://arxiv.org/abs/1702.04624}{	 arXiv:1702.04624}.

\bibitem{volov2} G. E. Volovik, {\it Flat band in the core of topological defects: bulk-vortex correspondence in topological superfluids with Fermi points}, \href{https://link.springer.com/article/10.1134/S0021364011020147}{JETP Lett. \textbf{93}, 66 (2011)}.

\bibitem{Mourik} V. Mourik, K. Zuo, S. M. Frolov, S. R. Plissard, E. Bakkers, L. P. Kouwenhoven, {\it Signatures of Majorana fermions in hybrid superconductor-semiconductor nanowire devices}, \href{http://science.sciencemag.org/content/336/6084/1003}{Science \textbf{336}, 1003 (2012)}.

\bibitem{Chang} W. Chang, V. E. Manucharyan, T. S. Jespersen, J. Nyg{\aa}rd, C. M. Marcus, {\it Tunneling Spectroscopy of Quasiparticle Bound States in a Spinful Josephson Junction}, \href{https://journals.aps.org/prl/abstract/10.1103/PhysRevLett.110.217005}{Phys. Rev. Lett. \textbf{110}, 217005 (2013)}.

\bibitem{Kjaergaard} M. Kjaergaard, F. Nichele, H. J. Suominen, M. P. Nowak, M. Wimmer, A. R. Akhmerov, J. A. Folk, K. Flensberg, J. Shabani, C. J. Palmstrom, C. M. Marcus, {\it Quantized conductance doubling and hard gap in a two-dimensional semiconductor-superconductor heterostructure}, \href{https://www.nature.com/articles/ncomms12841}{Nat. Commun. \textbf{7}, 12841 (2016)}.

\bibitem{Nichele} F. Nichele, A. Drachmann, A. M. Whiticar, E. O'Farrell, H. J. Suominen, A. Fornieri, T. Wang, G. Gardner, C. Thomas, A. T. Hatke, P. Krogstrup, M. J. Manfra, K. Flensberg, C. M. Marcus, {\it Scaling of Majorana Zero-Bias Conductance Peaks}, \href{https://journals.aps.org/prb/abstract/10.1103/PhysRevB.95.235305}{Phys. Rev. Lett. \textbf{119}, 136803 (2017)}.

\bibitem{Deng} M. T. Deng, S. Vaitiekenas, E. B. Hansen, J. Danon, M. Leijnse, K. Flensberg, J. Nyg{\aa}rd, P. Krogstrup, C. M. Marcus, {\it Majorana bound states in a coupled quantum-dot hybrid-nanowire system}, \href{http://science.sciencemag.org/content/354/6319/1557}{Science \textbf{354}, 1557 (2016)}.


\bibitem{Fu} L. Fu, C. L. Kane, {\it Superconducting proximity effect and Majorana fermions at the surface of a topological insulator}, \href{https://journals.aps.org/prl/abstract/10.1103/PhysRevLett.100.096407}{Phys. Rev. Lett. \textbf{100}, 096407 (2008)}.

\bibitem{Sau} J. Sau, R. Lutchyn, S. Tewari, S. Das Sarma, {\it Robustness of Majorana fermions in 2D topological superconductors}, \href{https://journals.aps.org/prb/abstract/10.1103/PhysRevB.82.094522}{Phys. Rev. B \textbf{82}, 094522 (2010)}.

\bibitem{Lutchyn} R. Lutchyn, J. Sau, S. Das Sarma, {\it Majorana Fermions and a Topological Phase Transition in Semiconductor - Superconductor Heterostructures}, \href{https://journals.aps.org/prl/abstract/10.1103/PhysRevLett.105.077001}{Phys. Rev. Lett. \textbf{105}, 077001 (2010)}.

\bibitem{dsarma} E. Dumitrescu, B. Roberts, S. Tewari, J. D. Sau, and S. Das Sarma, {\it Majorana fermions in chiral topological ferromagnetic nanowires}, \href{https://journals.aps.org/prb/abstract/10.1103/PhysRevB.91.094505}{Phys. Rev. B {\bf 91}, 094505 (2015)}.

\bibitem{allgpeople} R. Lutchyn, E. Bakkers, L. Kouwenhoven, P. Krogstrup, C. Marcus, Y. Oreg, {\it Realizing Majorana zero modes in superconductor-semiconductor heterostructures}, \href{http://xxx.lanl.gov/abs/1707.04899}{arXiv:1707.04899}.

\bibitem{ramon} R. Aguado, {\it Majorana quasiparticles in condensed matter}, \href{https://www.sif.it/riviste/sif/ncr/econtents/2017/040/11/article/0}{La Rivista del Nuovo Cimento 40, 523 (2017)}.

\bibitem{Kasahara} Y. Kasahara, T. Ohnishi, Y. Mizukami, O. Tanaka, S. Ma, K. Sugii, N. Kurita, H. Tanaka, J. Nasu, Y. Motome, T. Shibauchi, Y. Matsuda, {\it Majorana quantization and half-integer thermal quantum Hall effect in a Kitaev spin liquid}, \href{https://arxiv.org/abs/1805.05022}{arXiv:1805.05022}.

\bibitem{Shapiro} D.S. Shapiro, D.E. Feldman, A.D. Mirlin, and
  A. Shnirman, {\it Thermoelectric transport in junctions of Majorana and Dirac channels},
  \href{https://journals.aps.org/prb/abstract/10.1103/PhysRevB.95.195425}{Phys. Rev. B
    {\bf 95}, 195425 (2017)}.

\bibitem{Banerjee} M. Banerjee, M. Heiblum, V. Umansky, D. E. Feldman, Y. Oreg, A. Stern, {\it Observation of half-integer thermal Hall conductance}, \href{https://arxiv.org/abs/1710.00492}{arXiv:1710.00492}.

\bibitem{Perge} S. Nadj-Perge, I. K. Drozdov, J. Li, H. Chen, S. Jeon, J. Seo, A. H. MacDonald, B. A. Bernevig, A. Yazdani, {\it Observation of Majorana Fermions in Ferromagnetic Atomic Chains on a Superconductor}, \href{http://science.sciencemag.org/content/346/6209/602}{Science \textbf{346}, 602 (2014)}.

\bibitem{Pawlak} R. Pawlak, M. Kisiel, J. Klinovaja, T. Meier, S. Kawai, T. Glatzel, D. Loss, and E. Meyer, {\it Probing Atomic Structure and Majorana Wavefunctions in Mono-Atomic Fe-chains on Superconducting Pb-Surface}, \href{https://www.nature.com/articles/npjqi201635}{npj Quantum Information \textbf{2}, 16035 (2016)}.

\bibitem{Ruby} M. Ruby, F. Pientka, Y. Peng, F. von Oppen, Benjamin W. Heinrich, and K. J. Franke, {\it End states and subgap structure in proximity-coupled chains of magnetic adatoms}, \href{https://journals.aps.org/prl/abstract/10.1103/PhysRevLett.115.197204}{Phys. Rev. Lett. \textbf{115}, 197204 (2015)}.


\bibitem{Menzel} M. Menzel, Y. Mokrousov, R. Wieser, J. E. Bickel, E. Vedmedenko, S. Blügel, S. Heinze, K. von Bergmann, A. Kubetzka, and R. Wiesendanger, {\it Information Transfer by Vector Spin Chirality in Finite Magnetic Chains}, \href{https://journals.aps.org/prl/abstract/10.1103/PhysRevLett.108.197204}{Phys. Rev. Lett. \textbf{108}, 197204 (2012)}.

\bibitem{Jeon} S. Jeon, Y. Xie, J. Li, Z. Wang, B. A. Bernevig, A. Yazdani, {\it Distinguishing a Majorana zero mode using spin resolved measurements}, \href{http://science.sciencemag.org/content/358/6364/772}{Science \textbf{358}, 772 (2017)}.

\bibitem{Carvalho} A. Carvalho, M. Wang, X. Zhu, A. Rodin, H. Su, and A. C. Neto, {\it Phosphorene: from theory to applications}, \href{https://www.nature.com/articles/natrevmats201661}{Nat. Rev. Materials \textbf{1}, Article number: 16061 (2016)}.

\bibitem{kim1} J. Kim, S. S. Baik, S. H. Ryu, Y. Sohn, S. Park, B.-G. Park, J. Denlinger, Y. Yi, H. J. Choi, and K. S. Kim, {\it Observation of tunable band gap and anisotropic Dirac semimetal state in black phosphorus}, \href{http://science.sciencemag.org/content/349/6249/723}{Science {\bf 349}, 723 (2015)}.

\bibitem{kim2} J. Kim, S. Baik, S. Jung, Y. Sohn, S. Ryu,
H. Choi, B. Yang, and K. Kim, {\it Two-Dimensional Dirac Fermions Protected by Space-Time Inversion Symmetry in Black Phosphorus}, \href{https://link.aps.org/doi/10.1103/PhysRevLett.119.226801}{Phys. Rev. Lett. {\bf 119}, 226801 (2017)}.

\bibitem{Shirotani} I. Shirotani, J. Mikami, T. Adachi, Y. Katayama, K. Tsuji, H. Kawamura, O. Shimomura, and T. Nakajima, {\it Phase transitions and superconductivity of black phosphorus and phosphorus-arsenic alloys at low temperatures and high pressures}, \href{https://journals.aps.org/prb/abstract/10.1103/PhysRevB.50.16274}{Phys. Rev. B \textbf{50}, 16274 (1994)}.


\bibitem{Guo} J. Guo, H. Wang, F. von Rohr, W. Yi, Y. Zhou, Z. Wang, S. Cai, S. Zhang, X. Li, Y. Li, J. Liu, K. Yang, A. Li, S. Jiang, Q. Wu, T. Xiang, R. J. Cava, and L. Sun, {\it Electron-hole balance and the anomalous pressure-dependent superconductivity in black phosphorus}, \href{https://journals.aps.org/prb/abstract/10.1103/PhysRevB.96.224513}{Phys. Rev. B \textbf{96}, 224513 (2017)}.

\bibitem{Livas} J. Flores-Livas, A. Sanna, A. Drozdov, L. Boeri, G. Profeta, M. Eremets, and S. Goedecker, {\it Interplay between structure and superconductivity: Metastable phases of phosphorus under pressure}, \href{https://journals.aps.org/prmaterials/abstract/10.1103/PhysRevMaterials.1.024802}{Phys. Rev. Materials \textbf{1}, 024802 (2017)}.

\bibitem{Q.Huang} G. Q. Huang, Z. W. Xing, and D. Y. Xing, {\it Prediction of superconductivity in Li-intercalated bilayer phosphorene}, \href{http://stacks.iop.org/1674-1056/25/i=2/a=027402}{Appl. Phys. Lett. \textbf{106}, 113107 (2015)}.

\bibitem{Jun-JieZhang} J. Zhang and S. Dong, {\it
Prediction of above 20 K superconductivity of blue phosphorus bilayer with metal intercalations}, \href{http://iopscience.iop.org/article/10.1088/2053-1583/3/3/035006/meta}{2D Mater. \textbf{3}, 035006 (2016)}.

\bibitem{YanqingFeng} Y. Feng, H. Sun, J. Sun, Z. Lu and Y. You, {\it Prediction of phonon-mediated superconductivity in hole-doped black phosphorus}, \href{http://iopscience.iop.org/article/10.1088/1361-648X/aa9a5f/meta}{J. Phys.: Condens. Matter \textbf{30} 015601 (2018)}.


\bibitem{R.Zhang} R. Zhang1, J. Waters, A.K. Geim and I.V. Grigorieva, {\it Intercalant-independent transition temperature in superconducting black phosphorus}, \href{https://www.nature.com/articles/ncomms15036}{Nat. Comm. \textbf{8}, Article number: 15036 (2017)}.

\bibitem{Yuan} H. Yuan, L. Deng, B. Lv, Z. Wu, Z. Yang, S. Li, S. Huyan, Y. Ni, J. Sun, F. Tian, D. Wang, H. Wang, S. Chen, Z. Ren, C. Chu, {\it Investigation on the reported superconductivity in intercalated black phosphorus}, \href{https://arxiv.org/abs/1712.06716}{arXiv:1712.06716}.

\bibitem{beenakker} C. W. J. Beenakker, {\it Colloquium: Andreev reflection and Klein tunneling in graphene}, \href{https://doi.org/10.1103/RevModPhys.80.1337}{Rev. Mod. Phys. {\bf 80}, 1337 (2008)}.

\bibitem{Rudenko} A. N. Rudenko and M. I. Katsnelson, {\it Quasiparticle band structure and tight-binding model for single-and bilayer black phosphorus}, \href{https://journals.aps.org/prb/abstract/10.1103/PhysRevB.89.201408}{Phys. Rev. B \textbf{89}, 201408(R) (2014)}.


\bibitem{Voon1} L. Voon, A. Lopez-Bezanilla, J. Wang, Y. Zhang, M. Willatzen, {\it Effective Hamiltonians for phosphorene and silicene}, \href{http://iopscience.iop.org/article/10.1088/1367-2630/17/2/025004/meta}{New J. Phys. \textbf{17}, 025004 (2015)}.

\bibitem{Voon2} L. Voon, J. Wang, Y. Zhang, and M. Willatzen, {\it Band parameters of phosphorene}, \href{10.1088/1742-6596/633/1/012042}{J. Phys.: Conf. Ser. \textbf{633}, 012042 (2015)}.

\bibitem{Pereira} J. M. Pereira Jr. and M. I. Katsnelson, {\it Landau levels of single layer and bilayer phosphorene}, \href{10.1103/PhysRevB.92.075437}{Phys. Rev. B \textbf{92}, 075437 (2015)}.

\bibitem{wei} Q. Wei and X. Peng, {\it Superior mechanical flexibility of phosphorene and few-layer black phosphorus}, \href{http://aip.scitation.org/doi/abs/10.1063/1.4885215}{Appl. Phys. Lett. 104, 251915 (2014)}.

\bibitem{zyuzin} A. Zyuzin, M. Alidoust, D. Loss, {\it Josephson
    junction through a disordered topological insulator with helical
    magnetization}, \href{https://journals.aps.org/prb/abstract/10.1103/PhysRevB.93.214502}{Phys. Rev. B {\bf 93}, 214502 (2016)}.

\bibitem{bobkova} I. V. Bobkova, A. M. Bobkov, A. A. Zyuzin,
  M. Alidoust, {\it Magnetoelectrics in disordered topological
    insulator Josephson junctions},
  \href{https://journals.aps.org/prb/abstract/10.1103/PhysRevB.94.134506}{Phys. Rev. B
    {\bf 94}, 134506 (2016)}.

\bibitem{alidoustphi0} M. Alidoust and H. Hamzehpour, {\it Spontaneous supercurrent and
$\varphi_0$ phase shift parallel to magnetized topological insulator interfaces}, \href{https://journals.aps.org/prb/abstract/10.1103/PhysRevB.96.165422}{Phys. Rev. B 96, 165422 (2017)}.

\bibitem{Krive} I. V. Krive, A. M. Kadigrobov, R. I. Shekhter, and
  M. Jonson, {\it Influence of the Rashba effect on the Josephson
    current through a superconductor/Luttinger liquid/superconductor
    tunnel junction}, \href{https://journals.aps.org/prb/abstract/10.1103/PhysRevB.71.214516}{Phys. Rev. B {\bf 71}, 214516 (2005)}.

\bibitem{herve} A. Assouline, C. Feuillet-Palma, N. Bergeal, T. Zhang, A. Mottaghizadeh, A. Zimmers, E. Lhuillier, M. Marangolo, M. Eddrief, P. Atkinson, M. Aprili, H. Aubin, {\it Spin-Orbit induced phase-shift in $\rm Bi_2Se_3$ Josephson junctions}, \href{https://arxiv.org/abs/1806.01406}{arXiv:1806.01406}.


\bibitem{roso} A. Reynoso, G. Usaj, C. Balseiro, D. Feinberg, M. Avignon, {\it Anomalous Josephson Current in Junctions with Spin
    Polarizing Quantum Point Contacts},
  \href{http://dx.doi.org/10.1103/PhysRevLett.101.107001}{Phys. Rev. Lett. {\bf 101},
    107001 (2008)}.


\bibitem{Szombati} D. B. Szombati, S. Nadj-Perge, D. Car, S. R. Plissard, E. P. A.
M. Bakkers, and L. P. Kouwenhoven, {\it Josephson $\varphi_0$-junction in
  nanowire quantum dots}, \href{http://www.nature.com/nphys/journal/vaop/ncurrent/full/nphys3742.html?WT.feed_name=subjects_physics}{Nat. Phys. {\bf 12}, 568 (2016)}.


\bibitem{wylsuperc_exp1} C. Heikes, I.-L. Liu, T. Metz, C. Eckberg, P. Neves, Y. Wu, L. Hung, P. Piccoli, H. Cao, J. Leao, J. Paglione, T. Yildirim, N. P. Butch, W. Ratcliff, {\it Mechanical control of crystal symmetry and superconductivity in Weyl semimetal $\rm MoTe_2$}, \href{https://arxiv.org/abs/1804.09093}{arXiv:1804.09093}.


\bibitem{Marra1} P. Marra, R. Citro, C. Ortix, {\it Controlling Majorana states in topologically inhomogeneous superconductors}, \href{https://journals.aps.org/prb/abstract/10.1103/PhysRevB.95.140504}{Phys. Rev. B 95, 140504 (2017)}.


\bibitem{Soluyanov_app} A. A. Soluyanov, D. Gresch, Z. Wang, Q. Wu, M. Troyer, X.
Dai, and B. A. Bernevig, {\it Type-II Weyl semimetals}, \href{https://www.nature.com/articles/nature15768}{Nature (London) \textbf{527}, 495 (2015)}.

\bibitem{Ruan_app}  J. Ruan, S.-K. Jian, H. Yao, H. Zhang, S.-C. Zhang, and D. Xing, {\it Symmetry-protected ideal Weyl semimetal in HgTe-class materials}
\href{https://www.nature.com/articles/ncomms11136}{Nat. Commun. \textbf{7}, 11136 (2016)}.




\end{thebibliography}
\end{document}